\documentclass[%
 reprint,
superscriptaddress,
showpacs,
 amsmath,amssymb,
 amsfonts, 
 aps,
prb,
floatfix,
]{revtex4-1}

\pdfoutput=1

\usepackage{float}
 
\usepackage{graphicx}
\usepackage{dcolumn}
\usepackage{bm}

\usepackage{array}
\usepackage{makecell}
\usepackage{multirow}
\usepackage{adjustbox}
\newcolumntype{K}[1]{>{\centering\arraybackslash}p{#1}}
\newcolumntype{D}[1]{>{\arraybackslash}p{#1}}

\usepackage{xcolor}

\usepackage[normalem]{ulem}
\usepackage{cancel}

\begin{document}

\title{Edge effect on the current-temperature characteristic of finite area thermionic cathodes}

\author{Anna Sitek}
 \email{anna.sitek@pwr.edu.pl}
 \affiliation{Department of Engineering, Reykjavik University, 
	          Menntavegur 1, IS-102 Reykjavik, Iceland}
 \affiliation{Department of Theoretical Physics, Wroclaw University of Science and Technology, 
              Wybrze\.{z}e  Wyspia\'{n}skiego  27, 50-370 Wroclaw, Poland}

\author{Kristinn Torfason}
 \affiliation{Department of Engineering, Reykjavik University, 
  	          Menntavegur 1, IS-102 Reykjavik, Iceland} 
 
\author{Andrei Manolescu}
\affiliation{Department of Engineering, Reykjavik University, 
	         Menntavegur 1, IS-102 Reykjavik, Iceland}

\author{\'Ag\'ust Valfells}
 \affiliation{Department of Engineering, Reykjavik University, 
              Menntavegur 1, IS-102 Reykjavik, Iceland}

\begin{abstract} 
We perform a computational study, based on the molecular dynamics method, of the shape of Miram curves obtained from microscale planar diodes. We discuss the smooth transition from the source-limited to space-charge-limited regime due to the finite size of the emitter, i.e. the "knee" in the Miram curve. In our model we find that the smoothing occurs mostly due to the increased emission at the external edges of the emitting area, and that the knee becomes softer when the size of the emitting area decreases. We relate this to the recent work which has described how a heterogeneous work function similarly affects the Miram curve.
\end{abstract}


\maketitle

\section{\label{sec:introduction} Introduction}

The Miram curve is a characteristic of thermionic diodes showing the current dependence on the cathode temperature \cite{Cattelino82} (i.e., the $I$-$T$ characteristic). In the low-temperature range the current density is source limited or temperature limited and is given by the Richardson-Laue-Dushman (RLD) law \cite{Richardson16,Dushman23,Modinos84,Forbes20} and depends on the emitter work function and temperature. The number of particles in the diode gap increases with temperature, and thus also on the space-charge effects, which reduce the electric field at the cathode surface. At high temperatures the diode operates, in the space-charge-limited regime, in which a fraction of the released electrons is repelled back toward the cathode and the emission of other electrons is prevented. In this regime the current density depends on the applied voltage and gap spacing and is described by the Child-Langmuir law \cite{Child11,Langmuir23}. 

The transition region from the source-limited regime to the space-charge-limited regime is referred to as the "knee" in the Miram curve. This knee is particularly important because most thermionic diodes operate in the transition range \cite{Gilmour11} so as to generate large current density with minimal thermal degradation of the cathode. At the same time, this part of the Miram curve is the most difficult to understand. For the ideal model of an infinite, planar, and uniform cathode, the transition is very sharp and is followed by a stable (temperature-independent) current in the space-charge-limited regime \cite{Cattelino82,Miram04,Chernin20,Darr20}. In contrast, experimental Miram curves show smooth knees associated with cathode imperfections or disorder that translate into a nonuniform work function \cite{Cattelino82,Cattelino97,Wang09,Liu11}. 

Despite considerable effort to explain the shape of Miram curves and related temperature effects \cite{Cattelino82,Miram04,Cattelino82,Cattelino97,Gilmour11,Wang09,Liu11,Longo03,Vaughan86,Grant86,Chernin20,Darr20,Jassem21,Zhang17b}, the physical processes governing the thermal emission are not fully understood yet. Recently,  significant advancement has been made in understanding the processes responsible for the softening of the knee due to the studies of Chernin \textit{et al.} \cite{Chernin20} and Jassem \textit{et al.} \cite{Jassem21}. They investigated the impact of a nonuniform work function on the Miram curves obtained from infinite cathodes, and in particular they studied systems where the areas of different work functions formed parallel stripes, formed a checkerboard pattern, or were randomly arranged. Their findings showed that the inhomogeneity of the work function leads to a considerable softening of the knee, which increases with the size of the work-function grid. Chernin \textit{et al.} \cite{Chernin20} and Jassem \textit{et al.} \cite{Jassem21} associated the softening effect with the increased emission at the boundaries between regions of higher and lower work function. 

Recently we investigated field-assisted thermionic emission from cathodes with finite-size emitting areas \cite{Sitek21}. We observed the characteristic softening of the knee of the Miram curve and internal current distribution as Chernin \textit{et al.} \cite{Chernin20} and Jassem \textit{et al.} \cite{Jassem21} for cathodes with a heterogeneous work function, but we also observed that the knee of the Miram curve was softened even if the work function was homogeneous. We speculated that this was due to the finite size of the cathode and related to enhanced emission from the edges of the emitter.

In this work we apply the method used in Ref.\ \onlinecite{Sitek21} to further investigate the mechanisms responsible for softening of the knee region of the Miram curve. We look at the effect of the size of a finite emitter area of a uniform work function on the Miram curve, and compare it with the effect of work-function heterogeneity.


This paper is organized as follows. In the next section we describe the systems under study and the computational method. In Sec.\ \ref{sec:results} we present our results. 
Finally, Sec.\ \ref{sec:conclusions} contains a summary and final remarks.

\section{\label{sec:model} Model}

We study vacuum diodes consisting of infinite planar electrodes. The emitting area on the cathode surface is restricted to a square of variable side length ($L$) and is characterized by a uniform work function. Since the active region is embedded in the elsewhere nonemitting cathode, there is no field enhancement at the corners of the emitting area. The infinite lateral extent of the anode allows the absorption of every electron reaching its surface irrespective of the emitter size and the lateral expansion of the electron beam.    

The systems under study operate in the field-assisted-thermal-emission regime (i.e., the electrons are released thermally from the cathode surface over the top of a barrier reduced by the Schottky
effect but their movement within the gap is governed by the voltage applied between the anode and the cathode). In particular, for time $t < 0$, no electron is released,  due to a negative voltage bias that suppresses any current in the diode. At $t = 0$ this voltage is reversed and set to a finite value that allows the electrons to travel toward the anode. In this regime the voltage is too weak to release electrons from the cathode itself (i.e., it does not induce field emission). Nonetheless, it reduces the surface barrier via the Schottky effect and the space charge by sweeping the carriers away from the cathode. Both these effects result in an increase of the current with the applied voltage \cite{Sitek21}.        

To describe the current density injected into the system we use the formula derived in Refs.\ \onlinecite{Kevin,Kevin-tutorial}: 
\begin{eqnarray}
\label{func_J}	
 J(E,T) = A_{\mathrm{RLD}}T^2 n \int\limits^{\infty}_{-\infty}\textsl{}
\frac{\ln\left[1+e^{n\left(x-s\right)}\right]}{1+e^x}dx \, , 
\end{eqnarray}   
where $E$ is the electrostatic field at a point of the cathode surface (it is the result of the interplay between the applied voltage and space charge due to electrons filling the gap), $T$ represents temperature,  
\begin{eqnarray*}
	A_{\mathrm{RLD}} = \frac{qmk_{\mathrm{B}}^2}{2\pi^2\hbar^3}
\end{eqnarray*} 
is the Richardson constant, where $q$ and $m$ are the electron charge and mass, respectively, and $k_{\mathrm{B}}$ and $\hbar$ stand for the Boltzmann constant and the Planck constant, respectively, and 
\begin{eqnarray*}
	n = \frac{\beta_T}{\beta_E}  
\end{eqnarray*} 
is the ratio between the temperature and field-energy slope factors $\beta_T=1/{k_{\mathrm{B}}T}$ and $\beta_E$, respectively. Equation (\ref{func_J}) may be used to describe field and thermal emissions, as well as the transition range between these two processes. However, each regime requires particular forms of $\beta_E$ and the function $s$. According to Jensen \cite{Kevin}, for the case of field-assisted thermal emission 
\begin{eqnarray*}
   \beta_E = \frac{\pi}{q\hbar E} \sqrt{m\Phi\sqrt{\mathit{l}}} \, ,
\end{eqnarray*}	
where $\Phi$ represents the work function and
\begin{eqnarray*}
 \mathit{l}=\frac{q^3}{4\pi\epsilon_0}\frac{E}{\Phi^2}   
\end{eqnarray*}	
is a dimensionless positive parameter, while 
\begin{eqnarray*}
 s = \beta_E \phi \, , 
\end{eqnarray*} 
with $\phi = (1-\sqrt{l})\Phi$ the image-charge-reduced barrier.

After specifying the current density injected into the system [Eq.\ (\ref{func_J})], we apply the Metropolis-Hastings algorithm to determine the electric field at the cathode surface, and thus find favorable locations for the electrons to be emitted. Finally, we perform high-resolution molecular-dynamics simulations that include full Coulomb interactions of electrons. The method allows us to specify not only the spread and density of the whole beam in the gap but also the spread and density of beams originating from particular areas of the cathode, and thus determine currents (including partial currents) and beam quality factors \cite{Sitek21,Torfason21}.

We use the Ramo-Shockley theorem \cite{Ramo39} to calculate the currents,
\begin{equation*}
  I = \frac{q}{d}\sum_i \left(v_{z}\right)_i \ ,
\end{equation*}
where $d$ denotes the gap separating the anode from the cathode, $v_z$ stands for the $z$ component (normal to the cathode) of the electron's instantaneous velocity, and the summation is performed over all electrons in the gap. The Miram curves we present show the dependence of an averaged steady-state current on temperature. For each value of temperature we perform the full evolution of the system. 

We present the results of simulations performed for vacuum diodes with square emitting areas of side length $L$ ranging from $0.5$ to $6$ $\mu$m and constant gap spacing $d=1$ $\mu$m. We show Miram curves of uniform cathodes with work function $\Phi$ set to $2$ eV that are subjected to an applied voltage of $V_0=5$ V.

For such small cathodes the space-charge limit is reached at temperatures as high as 3000 K, and thus to 
take into account the space-charge-limited regime we need to consider temperatures beyond the realistic operating range of actual cathodes. We believe this is a scaling of parameters that does not change the physical phenomena that occur in realistic cathodes.  A similar extrapolation was used in Ref.\ \onlinecite{Sitek21}.

\section{\label{sec:results} Results}

Below we present our results on the Miram curves obtained from square emitters of finite size.
Because of the different current density between the central part of a cathode and the edge area, we treat these regions separately and show how each of them contributes to the total current. We then compare the Miram curves obtained from emitting areas of different size. Finally, we comment on how a nonuniform work function affects the Miram curves.

\begin{figure}[t]
	\centering
	\includegraphics[scale=0.56]{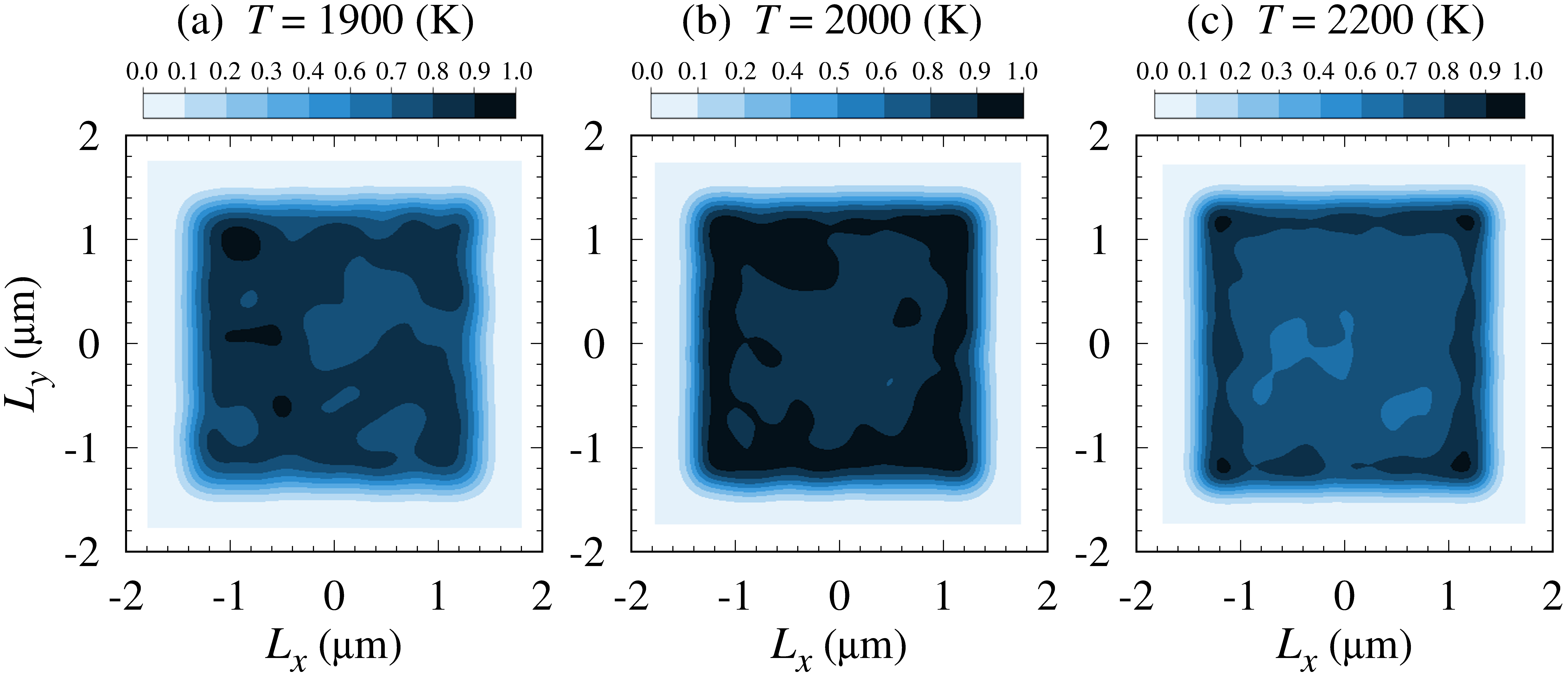}
	\caption{Normalized local density of electrons at the cathode with a uniform square emitting area of side length $L = 3$ $\mu$m for the transition (knee) region. 
	The color maps in each panel are normalized with respect to the largest value.}
	\label{fig:C_Map}
\end{figure}

A uniform thermionic cathode subjected to sufficiently high temperature and a bias voltage that allows the electrons to move toward the anode begins to emit electrons. Initially (at $T=1500$~K), only a few particles, which are released from random spots on the emitting surface, enter the gap. The number of emitted electrons increases with temperature, and at $T=1900$~K the local density of electrons at the cathode still has spatial randomness, as shown in Fig.\ \ref{fig:C_Map}(a). Nevertheless, all particles that enter the gap reach the anode. In this source-limited or temperature-limited regime, the emission, and thus the current, is governed by the properties of the cathode (e.g., work function) and temperature.

Further increase of temperature results in the accumulation of space charge in the gap, so electrostatic forces start to strongly affect the electron beam. At the cathode the current density is high at the edge of the emitter \cite{Umstattd01}, while in the interior region it is lower and, in the case of a homogeneous work function, uniform. This distribution is characteristic of space-charge-limited emission. As the beam propagates, space-charge forces cause the transverse beam envelope to grow and become rounded, and the current density to have a maximum at the center of the beam \cite{Sitek21}. As the current increases with temperature, and the beam is pushed into the space-charge-limited regime, edge emission becomes more enhanced as can be seen in Fig.\ \ref{fig:C_Map}(b) and Fig.\ \ref{fig:C_Map}(c). 
Moreover, the area of enhanced emission for temperatures above the knee is very narrow [Fig.\ \ref{fig:C_Map}(c)]. Recent work has shown increased emission within a very thin area around the boundary from circular emitters operating in the space-charge-limited regime, which is the source of the bulk of the current from microscopic finite emitters \cite{Gunnarsson21}.

\begin{figure}[t]
	\centering
	\includegraphics[scale=0.67]{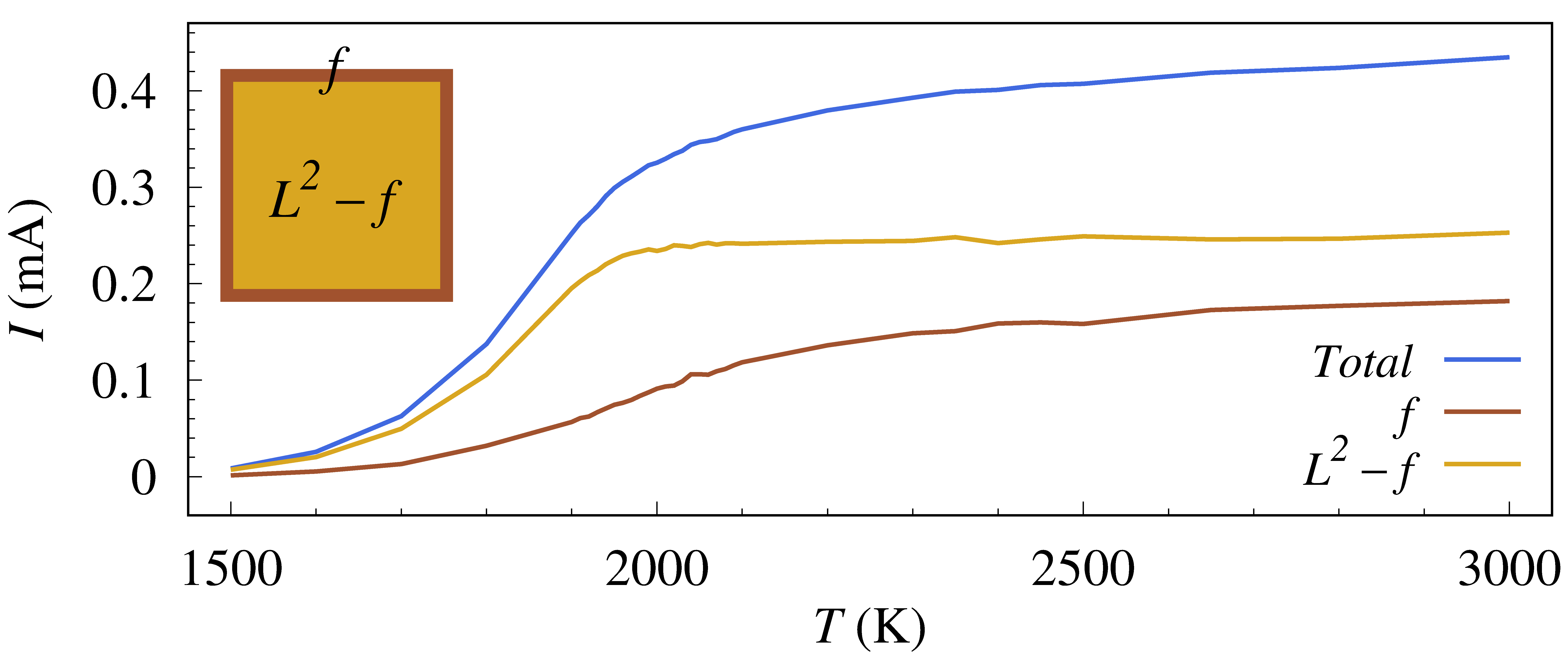}  
	\caption{Miram curve (blue) for a square emitting area of side length $L=3$ $\mu$m and its contributions from the "frame" region (brown) around the edges and from the remaining internal part (yellow) of the emitting area. The inset shows the emitting area divided into the frame region (brown) of width $t = 0.167$ $\mu$m and the internal area (yellow) for which the partial currents are calculated.} 
	\label{fig:Miram_Partial}	
\end{figure}

The current from the uniform cathode increases with temperature throughout the range studied (Fig.\ \ref{fig:Miram_Partial}, blue curve), but the rate of increase depends on the emission regime, and thus on temperature. At low temperatures (i.e., when the diode operates in the source-limited regime), the slope of the $I$-$T$ curve increases rapidly. During transition into the space-charge-limited regime the gradient decreases leading to the "knee" of the Miram curve. The total current never fully stabilizes, but increases with temperature even in the space-charge-limited regime.

The Miram curve for the finite diode with a homogeneous work function is thus qualitatively different from that of an infinite planar diode with a homogeneous work function. The latter has an abrupt transition between the source-limited regime and the space-charge-limited regime, as demonstrated in simulations 
based on solving the Vlasov and Poisson equations for a continuous charge distribution \cite{Chernin20,Jassem21}. 
It is of interest to note that the Miram curve for the finite-area emitter, the blue curve in Fig.\ \ref{fig:Miram_Partial}, not only shows the soft transition between source-limited and space-charge-limited emission, but also a continuing increase in emission with temperature in the space-charge-limited regime. Chernin \textit{et al.} observed this type of steady increase of current with temperature in particle-in-cell simulations for an infinite cathode of uniform work function \cite{Chernin20}, which they correctly attributed to the effects of finite emission velocity. In our model the electrons have vanishing emission velocity, so the increasing current in the space-charge regime is likely due to the finite area of emission, as demonstrated henceforth.

To better understand the shape of the Miram curve in Fig.\ \ref{fig:Miram_Partial}, we separate the contribution to the current from the emitter edge from that of the emitter interior by dividing the emitting area into the frame region ($f$) at the perimeter of the emitting area of the cathode and the remaining internal part, as shown in the inset in Fig.\ \ref{fig:Miram_Partial}. Next we calculate separately the currents originating from both regions.
We see that the current coming from the interior of the cathode, represented by the yellow curve in Fig.\ \ref{fig:Miram_Partial}, resembles the Miram curves achieved with models of infinite emitting area and uniform work function \cite{Cattelino82,Miram04,Chernin20,Darr20} (i.e.,  the current rapidly increases with temperature, and then it stabilizes at a nearly constant value).
This constant current is our point of reference in defining the width of the frame area because it confirms that the whole internal area is subjected to space charge. Thus, we continue to decrease the thickness of the outer region as long as we can obtain nearly stable current from the central part.
The knee for current coming from the interior is softer than predicted by one-dimensional beam models \cite{Cattelino82,Miram04,Darr20}. In contrast to infinite systems, where the space-charge effects are spatially homogeneous, in finite systems the area of space-charge-limited emission first appears in the central part and extends with increasing temperature toward the edges. Thus, in the transition range the emission from the internal region is not uniform; in particular, it is stronger in the outer part [Figs.\ \ref{fig:C_Map}(b) and\ \ref{fig:C_Map}(c)], which results in the softening of the transition range, but still this knee is much sharper than that of the total current.

In contrast, the edge current represented by the brown curve in Fig.\ \ref{fig:Miram_Partial} continues to grow with the temperature throughout the whole temperature range studied. This is in accordance with the well-known effect that in the space-charge limit the current density from the edge of a finite emitter is significantly larger than that from the interior \cite{Umstattd01,Gunnarsson21}. 
At low temperatures the whole emitter operates in the temperature-limited regime, and thus the magnitudes of partial currents are roughly proportional to the fractions of the area from which 
they originate.  This changes with increasing temperature, due to the space-charge effects, which are much greater in the central section than in the edge area [Figs.\ \ref{fig:C_Map}(b) and\ \ref{fig:C_Map}(c)]. For example, at $T=3000$ K the edge current exceeds $40\%$ of the total current although the area of the frame region constitutes only $21\%$ of the whole emitting area.

\begin{figure}[t]
	\centering
	\includegraphics[scale=0.67]{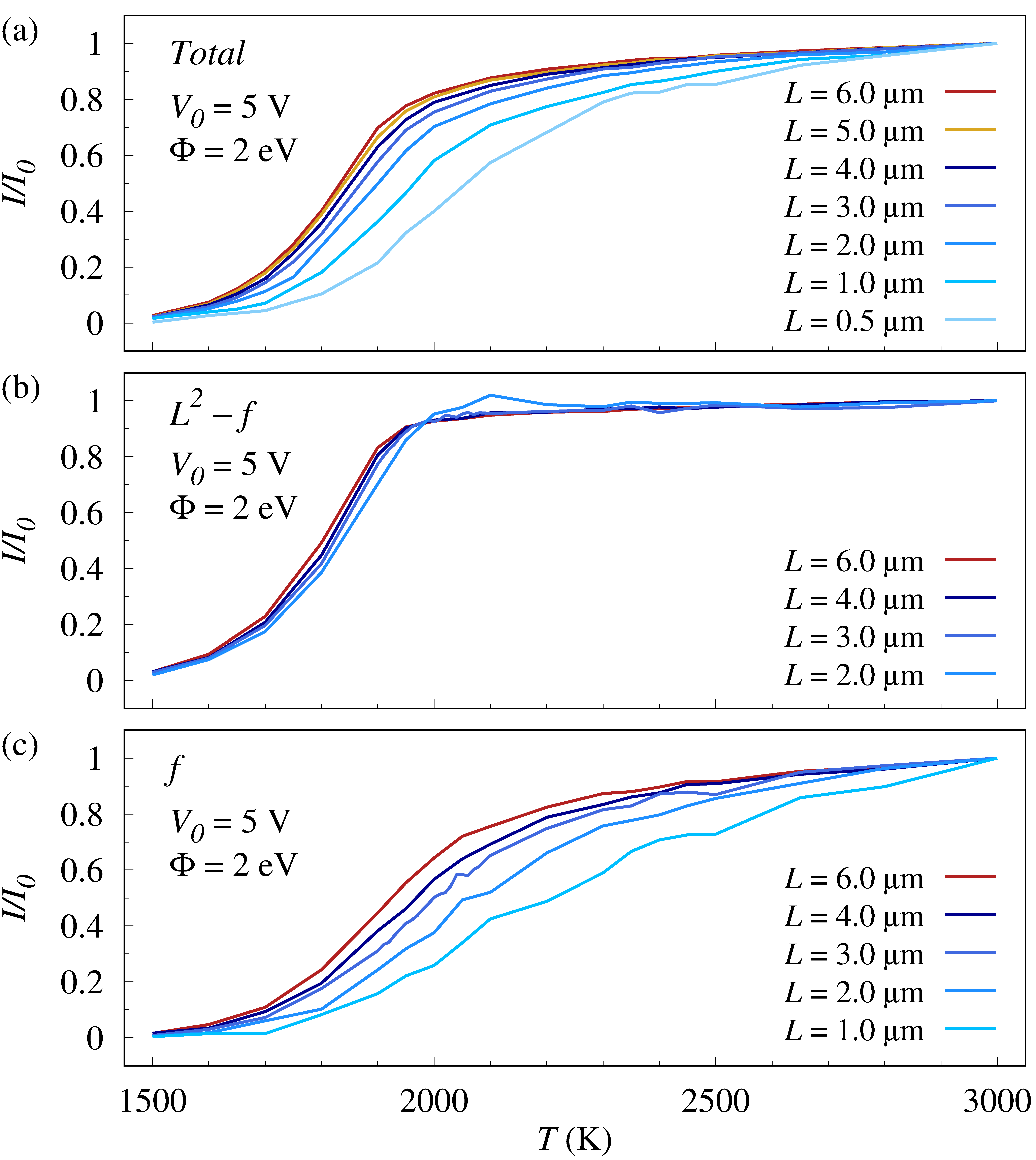}  
	\caption{Normalized Miram curves [$I_0 = I(T=3000$ K$)$] for cathodes of different side length $L$ (a) and the contributions from the internal region (b) and the perimeter region representing 21\% of the emitter area (c).} 
	\label{fig:Miram_L}	
\end{figure}

In our model the softening of the knee also depends on the size of the emitting area. In particular, the smaller the active region is, the smoother the transition from the source-limited regime to the space-charge-limited regime is [Fig.\ \ref{fig:Miram_L}(a)]. For a square of edge length $L$ the ratio between its perimeter length and area is $4/L$. This ratio increases as the edge length decreases, which 
implies an increased relative contribution from the edge on the emission from smaller emitters. 
In Figs.\ \ref{fig:Miram_L}(b) and\ \ref{fig:Miram_L}(c) we compare the partial currents obtained from the internal and frame regions, respectively. In all cases the area of the frame constitutes $21\%$ of the emitter area, while the internal part represents $79\%$. The shape of the curve representing the current originating from the central region does not depend on the edge length 
which is not surprising if we keep in mind that the space-charge emission limit at any point is primarily due to the electrons closest to that point. Thus, because of local homogeneity of emission, the $I$-$T$ curve for the interior resembles more that of an infinite emitter of uniform work function. At the emitter edge two things influence the space-charge limit. First, the emission outside the edge region decreases the space charge to one side and allows a higher current density in the edge region under space-charge-limited conditions. Second, the absence of space charge outside the edge region allows electrons emitted from the edge to expand outward, further accommodating more current from the edge. This results in a much-softer transition in the $I$-$T$ curve for the frame area.

It is illustrative to compare our results on how a finite emitter area affects the Miram curve with those obtained by varying the work function on the surface of an infinite emitter \cite{Chernin20, Jassem21, Sitek21}. In the infinite-area emitter the heterogeneous internal structure leads to a softening of the Miram curve due to piecewise space-charge limitation across the cathode surface. This piecewise limitation occurs because the current from the high-work-function regions reaches the space-charge limit at higher temperature and because electrons from the edge of the lower-work-function regions expand laterally into the area above the higher-work-function areas. It is the lateral expansion and discontinuity of space charge that is similar to the physics behind the effects of a finite emission area on the Miram curve. In the heterogeneous cathode the lateral expansion is internal and results in a uniform current density above the cathode surface, whereas in the finite emitter case, the current density is always greater at the edge.

\section{\label{sec:conclusions} Conclusions}

We study the field-assisted thermionic emission from planar cathodes with emitting areas of finite dimensions. We focus on  current versus temperature dependence (i.e. on the Miram curves). In particular, we investigate the transition range between source-limited emission and space-charge-limited emission commonly referred to as a "knee". According to many theoretical models this transition should be abrupt and followed by a space-charge-dominated constant current. On the contrary, the transition range of experimentally obtained characteristics is smoothly curved.   

Recent studies show that softening of the knee in the Miram curve may be attributed to a heterogeneous work function on the emitter surface \cite{Chernin20, Jassem21, Sitek21}. Our results show that the softening of the knee occurs also due to the finite size of the emitters. The electrons released close the edge of the emitting area may easily diffuse to the surrounding vacuum.
The emission at the edges is increased with respect to the central part of the cathode, and the corresponding edge current constitutes a considerable fraction of the total current and slowly increases with temperature. This slowly but constantly growing contribution is responsible for softening of the knee and for the increase of the total current in the space-charge-dominated regime. Moreover, we show that the degree of softening increases with decrease of the size of the emitting area.

\begin{acknowledgments}
This work was supported by the Air Force Office of Scientific Research under Grant No. FA9550-18-1-7011 and by the Icelandic Research Fund under Grants No. 174127 and No. 195943. 
Any opinions, findings, and conclusions or recommendations expressed in this material are those of the authors and do not necessarily reflect the views of the United States Air Force.
\end{acknowledgments}


%

\end{document}